\documentclass[12pt]{article}
\usepackage{setspace}
\usepackage{amsmath,amssymb,mathrsfs}
\usepackage{color}
\usepackage{hyperref}
\begin{document}
\setlength{\topmargin}{-1cm} 
\setlength{\oddsidemargin}{-0.25cm}
\setlength{\evensidemargin}{0cm}
\newcommand{\e}{\epsilon}
\newcommand{\beq}{\begin{equation}}
\newcommand{\eeq}[1]{\label{#1}\end{equation}}
\newcommand{\bea}{\begin{eqnarray}}
\newcommand{\eea}[1]{\label{#1}\end{eqnarray}}
\renewcommand{\Im}{{\rm Im}\,}
\renewcommand{\Re}{{\rm Re}\,}
\newcommand{\diag}{{\rm diag} \, }
\newcommand{\Tr}{{\rm Tr}\,}
\def\draftnote#1{{\color{red} #1}}
\def\bldraft#1{{\color{blue} #1}}
\def\n{n \cdot v}
\def\ni{n\cdot v_I}
\begin{titlepage}
\begin{center}

\vskip 4 cm

{\Large \bf Three Puzzles with Covariance and Supertranslation Invariance of Angular Momentum Flux (with Solutions)}

\vskip 1 cm

{Reza Javadinezhad \footnote{E-mail: \href{mailto:rj1154@nyu.edu}{rj1154@nyu.edu}} and Massimo Porrati \footnote{E-mail: \href{mailto:mp9@nyu.edu}{mp9@nyu.edu}} }

\vskip .75 cm

{\em Center for Cosmology and Particle Physics, \\ Department of Physics, New York University, \\ 726 Broadway, New York, NY 10003, USA}

\end{center}

\vskip 1.25 cm

\begin{abstract}
\noindent  
We describe and solve three puzzles arising in covariant and supertranslation-invariant formulas for the flux of angular momentum and other Lorentz charges in asymptotically flat spacetimes: 
1) Supertranslation-invariance and covariance imply invariance under spacetime translations; 2) 
the flux depends on redundant auxiliary degrees of freedom that cannot be set to zero in all Lorentz frames without  breaking Lorentz covariance;
   3) supertranslation-invariant Lorentz charges do not generate the 
transformations of the Bondi mass aspect implied by the isometries of the asymptotic metric.
    
 In this letter, we  solve the first two puzzles by presenting covariant formulas that unambiguously determine the auxiliary 
 degrees of freedom and clarify the last puzzle by explaining the different role played by covariant and canonical charges.  
 Our construction makes explicit the choice of reference frame underpinning seemingly unambiguous results 
 presented in the current literature.
 \end{abstract}
\end{titlepage}
\newpage

\section{Introduction}\label{intro}
The problem of a proper definition of angular momentum and boost charges for asymptotically flat spacetimes in general relativity has
 existed since it was discovered in~\cite{bond1,bms,s1,s2} that the metrics of spacetimes allowing for gravitational radiation
  enjoy an infinite dimensional asymptotic symmetry algebra. This is the BMS (Bondi-Metzner Sachs) algebra, which contains the 
  Poincar\'e algebra as a proper {\em but not normal} subalgebra. The Poisson brackets of angular momentum with supertranslations, which are an infinite-dimensional Abelian 
 subalgebra of BMS, do not vanish; consequently the total angular momentum and the flux of angular momentum  
  through null infinity can be changed by a gravitational wave of infinite wavelength. Since such a wave cannot be detected by any finite-size observer, the ambiguity due to BMS seems to preclude a 
 meaningful definition of angular momentum and the other Lorentz charges in general relativity (GR). Problems inherent in 
 defining angular momentum in GR were noticed by R. Penrose as early as 1964~\cite{pen}.
 
The problem is even sharper in quantum gravity because it implies that  two states differing by a supertranslation have the same energy, so in particular the vacuum state is infinitely degenerate. This can be seen as follows: denote schematically with $S^a$ the generators of 
 supertranslations and with $J^{\mu\nu}$ ($\mu,\nu=0,1,2,3$) the Lorentz generators, then  the commutator 
 $[S^a,J^{\mu\nu}]$ is nonzero so two zero-energy states differing by a supertranslation have different angular momenta. Explicitly,
 \beq
 J^{ij}|0\rangle=0 \Rightarrow J^{ij} [1+ c_aS^a]|0\rangle =[J^{ij},S^a]|0\rangle \neq 0, \qquad i,j=1,2,3,
 \eeq{int1}
 so $|0\rangle$ and $[1+ c_aS^a]|0\rangle$ are different (vacuum) states. In well-defined quantum field theories in flat spacetime and in holographic theories of quantum gravity in Anti de Sitter (AdS)
 space, instead, the vacuum is unique. The apparent non-uniqueness of vacuum in quantum field theories with 
 spontaneously 
 broken symmetries is resolved because the physical Hilbert space decomposes into {\em superselection sectors}~\cite{stro}
 while an infinity of states at zero energy is excluded in holhographic theories in AdS because the dual CFT has a unique 
 vacuum. Besides, infinite vacuum degeneracy would violate generalizations of the Bekenstein bound~\cite{bek} 
 such as~\cite{bbound}. 
 The previous observations suggest that the  $J^{\mu\nu}$'s appearing in the BMS algebra may not the correct operators 
 to associate to angular momentum and boosts. Other, unambiguous quantities should be found.
 
Life is easy in Minkowski spacetime because the Lorentz generators can be defined as 
$J^{\mu\nu}= \int_\Sigma n_\rho T^{\rho [\mu} x^{\nu]}$, 
where $\Sigma$ is a complete Cauchy surface with timelike normal $n_\rho$. Unfortunately, this 
formula becomes ill-defined in
 general relativity and must be substituted by an integral over a 2-surface at infinity, but this is precisely what introduces
 supertranslation ambiguities.
  
 The problems we have 
 outlined are not merely formal. In numerical general relativity, ambiguities due to supertranslations have 
 been shown to affect the computation of waveforms~\cite{numrel1,numrel2} that are essential tools for detection and study 
 of black hole mergers. The danger of supertranslation-dependent quantities is that they mix the effect of unobservable
 background noise due to very long waves to the signal due to a given physical process (gravitational scattering, 
 black hole mergers etc.). This danger manifest both at the level of formal definitions and in dealing with numerical 
 simulations that by construction have finite resolution. Finding unambiguous conserved charges and fluxes
 in general relativity is therefore a serious problem, all the more urgent now after computational advances in general 
 relativity (see e.g.~\cite{break}) and observational discoveries~\cite{bhmerg}.
 
 In fact, the first direct detection of gravitational waves from black hole mergers~\cite{bhmerg} 
 renewed interest in the quest for a better definition of symmetries and conserved charges in asymptotically flat 4D 
 spacetime. Novel supertranslation-invariant definitions for angular momentum and angular momentum flux as well 
 as for charges and fluxes of the other Lorentz algebra generators have recently appeared in the literature. To our knowledge, the first supertranslation invariant formula for angular 
 momentum was given in ref.~\cite{comp}. In fact, ref.~\cite{comp} defines the {\em Bondi charge} for angular momentum at 
 any retarded time so it also provides a definition of flux. Refs.~\cite{yau21a,yau21b} give an independent definition of 
 supertranslation-invariant 
 Lorentz Bondi charges which, while agreeing with~\cite{comp} at retarded time $u\rightarrow -\infty$ differ at finite $u$. 
 Refs.~\cite{jkp,hanoi} provide formulas for supertranslation-invariant Lorentz charges and their fluxes based on a canonical 
 formalism that applies equally well  to the  quantum theory. 
 
The angular momentum flux given in~\cite{comp,jkp,hanoi} begins at $O(G^3)$ in a perturbative expansion in 
powers of the Newton constant $G$. This is in disagreement with explicit computation of mechanical angular momentum 
flux done by several groups with different methods. In particular both~\cite{bd} 
using techniques developed in~\cite{dam} and~\cite{conf1,conf2,conf3,conf4,conf5,man,heis} agree on the presence of a nonvanishing
angular momentum flux at $O(G^2)$. The origin of this discrepancy was traced back in~\cite{vv} to the difference between
 the supertranslation frame used in perturbative calculation of gravitational scattering and a ``canonical'' frame in which the 
 Bondi angular momentum at $u=-\infty$ agrees~\cite{ash} 
 with the ADM (Arnowitt, Deser, Misner) definition~\cite{adm}. Ref.~\cite{jp} employs the results of~\cite{vv}
to define, to all orders in $G$, a supertranslation-invariant angular momentum flux that agrees to $O(G^2)$ with perturbative 
calculations and is defined only in term of asymptotic metric data on $\mathscr{I}$.

All supertranslation-invariant formulas for the Lorentz charges flux depend on $C(u,\Theta)$, the ``electric'' 
component of the shear $C_{AB}(u,\Theta)$. These quantities are defined in the next section. The shear is 
independent of the first two harmonics of $C(u,\Theta) $ while the invariant flux depends on them. This
fact requires an independent choice of the $l=0,1$ harmonic components of $C(u,\Theta)\equiv \sum_{l=0}^\infty\sum_{-l\leq m \leq m} 
C_{l,m}(u)\mathrm{Y}_{lm}(\Theta)$. 
The simplest choice would be
to set them to zero {\em in all Lorentz frames}. Besides being aesthetically unpleasant, this choice is inconsistent with
Lorentz covariance because we will see that boosts mix the higher harmonics of $C$ with the $l=0,1$ ones. 
So, neither the Lorentz
charges flux in~\cite{yau21a,yau21b} nor those in~\cite{jp} are Lorentz covariant unless a new prescription is found. 
The effect of this non-invariance was explicitly
verified in the case of two-particle scattering in~\cite{rvw}, where it was shown that while the charge defined in~\cite{jp} agrees with 
perturbative calculations in the center of mass rest frame of the two particles, it does differs in the rest frame of one of the two
particles. 

In fact, all existing proposals for a supertranslation-invariant flux of Lorentz charges share an even more serious flaw: any 
covariant formula for the flux (or the Lorentz charge)  is a Lorentz tensor and no such quantity can be 
supertranslation invariant without being also invariant under spacetime {\em translations}. The proof follows simply from the 
structure of the BMS algebra and is valid also for the quantum BMS algebra.  So, whether we use the formulas 
of~\cite{comp,yau21a,yau21b,jkp} or~\cite{jp} one
thing is clear: whatever we are computing is an angular momentum {\em defined with respect to a particular choice of the 
origin of coordinates}. 

In this letter we show that the need to define an ``intrinsic'' angular momentum 
flux depending on an independently prescribed origin
of  coordinates is not a problem but rather the feature that allows us to solve the puzzles due to the ambiguity 
in the choice of $C_{lm}$, $l=0,1$ and to the lack of covariance. 
These are not mere technicalities but foundational 
problems that have so far precluded a covariant and unambiguous definition of angular momentum in general relativity and
have also introduced redundant quantities,  $C_{lm}$, $l=0,1$ without apparent physical meaning. 
Our paper instead shows
that they do have a clear meaning; we will show that they define the origin of the space-time coordinate system. 
Angular momentum depends on the point from which it is computed, but the quantities that determine that point
are nowhere to be found among the asymptotic components of the metric. 
We will show that the ``redundant'' s- and -p wave components of the electric shear are precisely those ``missing'' quantities.

We being  by defining our notations and deriving the key result that supertranslation-invariant definitions of
the flux must also be spacetime translation invariant.
Next we derives fully covariant formulas for the charges and fluxes defined in 
refs.~\cite{yau21a,yau21b}. Key to the covariantization is the use of the transformation law of the boundary graviton under boosts,
together with covariant and supertranslation-invariant definitions of the center of mass frame, which are used to determine the first two
harmonics of the boundary graviton.
 Finally, we give a physical interpretation to the
supertranslation-invariant charges we 
previously defined, shows that differences among different prescriptions do not manifest
themselves until $O(G^3)$, and shows how current calculations found in the literature implicitly use the same choice of 
reference frame as our own.

We also solve the third puzzle by clarifying the difference between supertranslation invariant charges and the
generators of asymptotic symmetries.

\section{Notations and transformation properties of covariant quantitites}\label{review}

The metric near future null infinity $\mathscr{I}^+$ in Bondi-Sachs coordinates is~\cite{bond1,bms,s1,s2}
\bea
 ds^2  &=&  -du^2 - 2 du\, dr + r^2 \left(h_{AB}
+\frac{C_{AB}}{r}\right) d\Theta^A d\Theta^B + D^AC_{AB}\, du\, d\Theta^B +
\frac{2m}{r} du^2  \nonumber \\ && 
		+ \frac{1}{16 r^2} C_{AB}C^{AB} du dr 
	 + \frac{1}{r}\left(
		\frac{4}{3}\left(N_A+u \partial_A m \right) - \frac{1}{8} \partial_A \left(C_{BD}C^{BD}\right)
		\right) du d\Theta^A \nonumber \\ &&
		+ \frac{1}{4} h_{AB} C_{CD}C^{CD}d \Theta^A d \Theta^B
		+ \dots , 
		\eea{1}
where the mass aspect $m(u,\Theta)$ is a scalar, the angular momentum aspect $N_A(u,\Theta)$ is a vector and the shear $C_{AB}(u,\Theta) $ is a symmetric traceless tensor. All these quantities are defined on 
the celestial sphere with coordinates $\Theta_A$ and round metric $h_{AB}$ and also depend on the retarded time $u$. 
The dots in~\eqref{1} denote subdominant terms in $1/r$. 
The coordinate system in eq.~\eqref{1} is invariant under the asymptotic symmetries $u\rightarrow u + f(\Theta)$, called  
supertranslations~\cite{bms}. Energy, momentum and Lorentz charges 
are defined in terms of $m$, $N_A$, the $l=1$ spherical harmonics $\vec{\mathrm{Y}}$, and the six conformal Killing 
vectors $Y^A$ of the celestial sphere~\cite{bms}
\bea
E(u)&=&{1\over 4\pi G} \int d^2 \Theta \sqrt{h} m(u,\Theta), \quad
\vec{P}(u)={1\over 4\pi G} \int d^2 \Theta \sqrt{h} \vec{\mathrm{Y}}m(u,\Theta), \nonumber \\ 
J_Y(u) &=&  {1\over 8\pi G} \int d^2 \Theta \sqrt{h} Y^A N_A(u,\Theta).
\eea{2}
By definition the conformal Killing vectors obey $D_A Y_B + D_B Y_A= h_{AB} D_CY^C $. For rotations $D_CY^C=0$ while for boosts
$D_CY^C$ is an $l=1$ harmonic. In the latter case we can write $Y_A=D_A\psi$ with $\psi$ obeying $D_AD^A \psi = -2 \psi$. 
When it is well defined, the  total energy flux $\Delta E \equiv E(+\infty)-E(-\infty)$ is invariant under supertranslations. This is not true 
for the flux of the Lorentz charges; so in particular the angular momentum flux
$\Delta J_Y \equiv J_Y(+\infty)-J_Y(-\infty)$ can be changed by a supertranslation. 
   
The Lorentz Bondi charges at retarded time $u$ defined in~\cite{yau21b} are 
\bea
J^{CWWY}_Y(u) &=& J_Y(u) -j_Y[m(u),C(u)]  \nonumber \\
j_Y[m(u),C(u)] &=& {1\over 8\pi G} \int d^2 \Theta \sqrt{h} Y^A [ 3m(u,\Theta) D_A C(u,\Theta) + D_A m(u,\Theta) C(u,\Theta) ] 
\nonumber \\ &=& {1\over 4\pi G} \int d^2 \Theta \sqrt{h} m (\delta^{-1/2}_Y C) =
 -{1\over 4\pi G} \int d^2 \Theta \sqrt{h} (\delta^{3/2}_Y m) C,
 \eea{1a}
 where  
 \beq
 \delta^{w}_Y F\equiv w D\cdot Y F + Y\cdot DF.
 \eeq{1b}
The ``electric shear'' $C(u,\Theta)$  is defined by 
\beq
D^AD^B C_{AB}(u,\Theta)= D^2(D^2+2) C(u,\Theta). 
\eeq{2a}
The operator $D^2(D^2+2)$ is diagonalized by spherical harmonics, on which 
$D^2(D^2+2) C_l= l(l^2-1)(l+2)C_l$, so the $l=0,1$ harmonics in $C$ do not appear at all in the asymptotic 
metric.
The constraint relating the change in the mass aspect to the change in the electric 
shear is
\bea
\Delta m(\Theta)  &=& \frac{1}{4}  D^2(D^2+2)\Delta C(\Theta)  
-\int_{-\infty}^{+\infty} du T_{uu}(u,\Theta) ,\label{5a}  \\
\Delta C(\Theta) &=& C(+\infty,\Theta) - C(-\infty,\Theta), \quad \Delta m(\Theta)= m(+\infty,\Theta) - m(-\infty,\Theta).
\eea{5}
Here $T_{uu}={1\over 8}N_{AB}N^{AB} + \lim_{r\rightarrow\infty} r^2 T^M_{uu},$ with $T^M=$matter stress-energy tensor and 
$N_{AB}\equiv \partial_u C_{AB}$ is the Bondi news. In gravitational scattering $N_{AB}=O(G^2)$ so, through $O(G^2)$, 
 $\Delta m(\Theta)  = \frac{1}{4}  D^2(D^2+2)\Delta C(\Theta)$. This equation shows a further ambiguity in $C_{l=0,1}(u,\Theta)$: their 
 dependence on the retarded time $u$ is completely arbitrary. They are undetermined even if the initial value for all harmonics of 
 $C$ is given at $u=-\infty$. 
 
 We now prove that a supertranslation-invariant tensor must also be spacetime translation invariant by computing the commutator
  \beq
 \delta_Y \delta_f \Delta J_Z^{CWWY} - \delta_f \delta_Y \Delta J_Z^{CWWY}  = \delta_{[Y,f]}  \Delta J_Z^{CWWY} ,
 \eeq{mprl3} 
 with $f$ a supertranslation and $Y$ a Lorentz boost.
 The transformation of the flux is determined by Lorentz covariance to be 
 $\delta_Y \Delta J_Z^{CWWY}=\Delta J_{[Y,Z]}^{CWWY}$ where $[Y,Z]\equiv \mathcal{L}_Y Z - \mathcal{L}_Z Y$ with $\mathcal{L}_W$
 the Lie derivative along the vector $W$. Invariance under supertranslations therefore implies that the left hand side of 
 eq.~\eqref{mprl3} vanishes. The right hand side is the transformation of the flux under $[Y,f] \equiv \delta_Y^{-1/2} f$. A boost can be written as $Y^A=D^A\psi$ with $\psi$ an $l=1$ harmonic obeying $D^2\psi =-2 \psi$. So, for an $l=2$ supertranslation obeying
 $D^2 f= -6 f$ we find
 \beq
 [Y,f]= -{1\over 2} f (D^2 \psi)  +{1\over 2} D^2 (f \psi ) + f\psi + 3 f\psi 
 \eeq{prlm4}
 The product $f\psi$ contains an $l=1$ harmonic on which $D^2(f\psi)|_{l=1}= -2 (f\psi)|_{l=1}$ therefore 
 $[Y,f]|_{l=1}= 4(f\psi)|_{l=1}\neq 0$ and the boost of a supertranslation is a nonzero spacetime translation. 
 
 The flux 
 $\Delta J_Y^{CWWY}= J_Y^{CWWY}(+\infty)-J_Y^{CWWY}(-\infty) $ is manifestly translation and supertranslation 
 invariant if we define the transformation law of $C(u,\Theta)$ to be
 \beq
 C(u,\Theta)\rightarrow C'(u,\Theta) =C(u,\Theta) + f(\Theta)
 \eeq{mprl2} 
 For {\em any} function $f(\Theta)$. Its $l=0,1$ harmonics are spacetime translations, so the $l=0,1$ harmonics of $C(-\infty,\Theta)$ 
 represent the choice of origin for the coordinate system used to define the angular momentum.
 
 The upshot of this analysis is that any supertranslation invariant flux is necessarily an intrinsic flux, defined with respect to an origin of
 the coordinate system, which is equivalent to a choice of $C|_l$ for $l=0,1$. {\em An independent, covariant 
 prescription for the $l=0,1$ components of the boundary graviton is necessary to define  the flux.}
 
 We cannot simply set $C|_{l\leq 1}=0$ in all Lorentz frames because this choice is inconsistent with Lorentz transformations. 
 The problem is that the $l$-th harmonic  of the boundary graviton transforms under boosts exactly like a supertranslation;
 therefore, in parallel with eq.~\eqref{prlm4} we find
 \beq
 \delta^{-1/2}_Y C = -{1\over 2} D\cdot Y C + Y\cdot D C= \psi C + D\psi \cdot DC = {1\over 2} \left[ D^2 +4 +l(l+1)\right](\psi C_l) .
 \eeq{15a}
 The same argument as given after eq~\eqref{prlm4} shows that  $\delta^{-1/2}_Y C $ generically contains a nonvanishing 
 $l=1$ harmonic.
 
\section{Lorentz covariant definitions of angular momentum and boost charges and fluxes}\label{cov}

We found in the previous section that an independent definition for $C(u)|_{l\leq 1}$ is necessary to completely define 
$\Delta J^{CWWY}_Y$. The prescription should maintain covariance and should not introduce an additional arbitrariness in the flux.
Here we propose a simple recipe, which is covariant by construction. We also present a small variation on the prescription
that we later prove to coincide with the previous one to $O(G^2)$.

Let us consider the flux, computed in the initial center of mass rest frame (CMRF), which is defined by the condition 
$m^-_{1,m}\equiv\int d^2\Theta \sqrt{h} \mathrm{Y}_{1m} m(-\infty,\Theta)=0$. The definition of the frame
 is not complete because the origin of the coordinate
system can be translated arbitrarily in space and time. We remove this arbitrariness by requiring that in the CMRF the initial boost
charge vanish. We denote by $\pm$ $u$-dependent quantities evaluated at $\pm \infty$ and impose
\beq
J_{\bar{Y}}^- - j_{\bar{Y}}[m^-,C^-]=0 , \quad\mbox{ for all } \bar{Y}^A=\mathrm{boost}=D^A \psi, \quad D^2\psi=-2 \psi.
\eeq{prlm5} 
These are three conditions that uniquely determine the three components of $C|_{l=1}$. Explicitly, we use eq.~\eqref{1a} and 
\bea
  \delta^{3/2}_Y m^-_l  &=& {3\over 2} D\cdot Y  m^-_l  + Y\cdot D  m^-_l = -3\psi  m^-_l + D\psi \cdot D m^-_l  = {1\over 2} \left[ D^2 -4 +l(l+1)\right](\psi m^-_l) , \nonumber \\  && (D^2 +2 )(\psi  m^-_2)  |_{l=1}=0
    \eea{prlm5a}
to obtain
\beq
j_{\bar{Y}}[m^-,C^-|_{l\leq 1}]= j_{\bar{Y}}[m^-|_{l\leq 1},C^-|_{l\leq 1}]= {3m_0^-\over 8\pi G} \int d^2\Theta \sqrt{h} D\psi \cdot D C^-|_{l\leq 1} = {3m_0^-\over 4\pi G} \int d^2\Theta \sqrt{h} \psi \cdot  C^-|_{l\leq 1}.
\eeq{prlm6}
Eq.~\eqref{prlm5} then reduces to
\beq
    {3m_0^-\over 4\pi G}C_{1m}^-= J^-_{\bar{Y}^A} -
    j_{\bar{Y}^A}[m^-,C^-|_{l>1}], \quad \bar{Y}^A=D^A \mathrm{Y}_{1-m}
    \eeq{prlm7}

Under {\em both}  supertranslations and spacetime translations and for any conformal Killing vector $Y$ 
$J^-_Y \rightarrow J^-_Y + {1\over 4\pi G} \int d^2\Theta \sqrt{h} m^- \delta^{-1/2}_Y f $ so $J_Y^- - j_Y[m^-,C^-]$ is
by construction invariant. Our definition specifies the origin of the system of coordinates (up to a time translation). We are free to set
 $C|_{l=0}$ to whichever value we want and we will choose $C|_{l=0}=0$.

To extend our definition to any Lorentz frame with celestial sphere coordinates $\Theta=(\theta,\phi)$ we choose three conformal 
Killing vectors $\hat{Y}^A$, related to the pure boosts of the rest frame (with coordinates $\bar{\Theta}=(\bar{\theta},\bar{\phi})$ by
\beq
\hat{Y}^A(\Theta)= {\partial g^A\over \partial \bar{\Theta}^B}\bar{Y}^B(\bar{\Theta}), \quad
 \bar{Y}_B = {\partial \bar{\psi} \over \partial  \bar{\Theta}^B},
 \eeq{prlm8}
 evaluated at $\Theta^A=g^A(\bar{\Theta})$. We fix the Lorentz transformation $g^A$  from the CMRF to the Lorentz frame moving with 
 velocity $\vec{\beta}$ by requiring that it is a pure boost along $\vec{\beta}$. Covariance is preserved by this choice as it is most easily seen by writing the Lorentz charge as an antisymmetric matrix $J$, the boost from the CMRF as a pseudo-orthogonal matrix $B(\vec{\beta})$ and the constraint as 
 \beq
  Jy=0, \quad y= B(\vec{\beta}) \bar{y}, \quad \bar{y}=\begin{pmatrix} 1 \\0 \\0 \\ 0 \end{pmatrix} .
 \eeq{prlm8a}
 This definition is covariant because any Lorentz transformation can be decomposed as $L=BR$, with $B$ a pure boost and $R$ a pure rotation. In the CMRF
 $R\bar{y}=\bar{y}$ so $L\bar{y}=B\bar{y}$.
 Definition~\eqref{prlm8a} is also unique because Lorentz transformation $L$ to a frame moving with velocity $\vec{\beta'}$ maps $y$ into 
 $y'= L B(\vec{\beta})\bar{y}$. This is not a pure boost from the CMRF but the difference is a pure rotation of the CMRF, 
 $B(-\vec{\beta'})LB(\vec{\beta})=R$, hence $LB(\vec{\beta})\bar{y} =B(\vec{\beta'})R  \bar{y}= B(\vec{\beta'})\bar{y}$.
 So, a translation, supertranslation and covariant definition of the Lorentz charges is
 \beq
 \mathfrak{J}_Y^-=J^-_Y -j_Y[m^-,C^-],
 \eeq{prlm9}
 where the $l=0,1$ harmonics of the boundary graviton are given by solving the equation
 \beq
 J^-_{\hat{Y}} -j_{\hat{Y}}[m^-,C^-] =0 .
 \eeq{prlm10}
 
 To define the flux we have several possibilities. For all of them
 \beq
 \mathfrak{J}_Y^+=J^+_Y -j_Y[m^+,C^+], \quad \Delta  \mathfrak{J}_Y = \mathfrak{J}_Y^+ -  \mathfrak{J}_Y^-. 
  \eeq{prlm11} 
  The difference is in the choice of the equations for $C|_{l\leq 1}$. Two simple choices are
 \begin{description}
 \item{A} Set $C^+|_{l\leq 1}= C^-|_{l\leq 1}$.
 \item{B} Solve $J^+_{\tilde{Y}} -j_{\tilde{Y}}[m^+,C^+] =0$, where $\tilde{Y}^A$ is defined by Lorentz-transforming pure boosts defined in
 the {\em final} center of mass rest frame.
 \end{description}
 We show in the next section that these two definitions coincide to $O(G^3)$.
 
 \section{Interpretation of the invariant charges and fluxes}\label{int}
 
 We proposed a Lorentz-covariant, supertranslation-invariant definition of the Lorentz Bondi charges at $u=\pm \infty$, 
 which  differs from~\cite{yau21a,yau21b} only in the condition used to fix the $l=0,1$ components of the boundary 
 graviton. 
By construction both $\mathfrak{J}^\pm_Y$ are invariant under supertranslations so we can compute them by
changing coordinates in the non-radiative far past and far future regions to set $C_{AB}^\pm=0$. 
\beq
\Delta \mathfrak{J}_Y = J_Y(+\infty)|_{C^+|_{l>1}=0} - J_Y(-\infty)|_{C^-|_{l>1}=0}.
\eeq{25}
So the covariant, supertranslation-invariant flux reduces to the difference of canonical Bondi charge at $+ \infty$, computed in
the frame where the angular metric at $u=+\infty$ is $h_{AB}+ O(1/r^2)$, minus the canonical Bondi charge at $u=-\infty$, 
computed in the frame where the angular metric at $u=+\infty$ is also $h_{AB}+ O(1/r^2)$.  Refs.~\cite{vv,ash} argue that 
$h_{AB}+ O(1/r^2)$ is the frame where the Bondi charge reduces to the ADM charge. This identification suggests a very
natural interpretation of $\Delta \mathfrak{J}_Y$: it is the canonical charge measured after a gravitational scattering  in a  ``round metric'' 
frame minus the initial canonical charge, also measured in a ``round metric'' frame. The reference frames are fixed by requiring that the 
initial metric is round before the scattering occurs, then after the scattering has occurred  by requiring that the 
final metric is also
round. This procedure ``forgets'' the initial frame-fixing.
The interpretation of the flux and its form is essentially the same as in~\cite{yau21a,rvw,mww} and~\cite{dvhrv} once the 
subtleties due to Lorentz covariance and covariant 
subtraction of  the low-$l$ harmonics in $C^\pm$ are properly taken into account using eq.~\eqref{prlm10} for 
$C^-|_{l\leq 1}$ and either prescription A or B for $C^+|_{l\leq 1}$.
The Lorentz covariant prescription~\eqref{prlm10} can be used also
to covariantize the flux defined in~\cite{jkp}. It differs from
$\Delta\mathfrak{J}_Y$ by terms proportional to the gravitational memory
$C^+-C^-$. This computation was performed in the CMRF
in~\cite{rvw,mww} but the role of gravitational memory in the definition of
angular momentum in general realtivity was noticed long before, e.g.
in~\cite{winicour}. We will expand on this question in a future
publication~\cite{jpfuture}.

The difference between prescription A and B can be seen most clearly for the flux of angular momentum $\Delta\vec{J}$. 
Prescription A computes
the difference between initial and final angular momenta in the initial CMRF. So it includes a term due to the motion of the
final CMRF, namely $\Delta \vec{J}=\Delta\vec{J}^{intrinsic} + \vec{a} \times \Delta\vec{P}$, with $\vec{a}$ the displacement 
of the origin of the final CMRF with respect to the initial CMRF. Prescription B instead gives 
$\Delta \vec{J}=\Delta\vec{J}^{intrinsic}$. The difference between the two prescriptions amounts to a term proportional
to $\Delta\vec{P}$, i.e. the change of the center of mass momentum due to gravitational radiation. Because of 
the constraint equation~\eqref{5a}, the definition of momentum~\eqref{2} and $T_{uu}=O(G^4)$, 
$\Delta\vec{P}=O(G^3)$ so prescriptions A and B agree to $O(G^2)$.

We defined the $l=0,1$ harmonics of $C$ by requiring that the boost charges vanish in the initial CMRF. This is the same
prescription used in e.g. ref.~\cite{man}, which considered the scattering of two particles of initial 4-momenta $p^\mu_i$,
$i=1,2$ and impact parameters $b_i$ obeying $p_i \cdot b_j$=0 for all $i,j$. In the CMRF these constraints are
satisfied by $p_1=(E_1, p,0,0)$, $p_2=(E_2,-p,0,0)$ so $b_1=(0,0,y_1,z_1)$, $b_2=(0,0,y_2,z_2)$ and the nonzero boost 
charges are $J^{02}=E_1y_1 +E_2y_2$, $J^{03}=E_1z_1 + E_2 z_2 $. Setting $J^{02}=J^{03}=0$ we reproduce the
{\em particular} solution of the constraints chosen in~\cite{man}. Similar choices are done in other papers on gravitational 
scattering. They are natural because they set the origin of coordinates at the center of mass of the incoming particles.

\section{Difference between two consistent definitions of Lorentz charges}\label{diff}
By construction {\em any} definition of  supertranslation-invariant Lorentz generators $J^{inv}_Y$, so in
particular $J^{inv}_Y=\mathfrak{J}_Y^-$,
makes them commute with the $l>1$ harmonics of the mass aspect
  $m^-(\Theta)$ 
 \beq
 [J^{inv}_Y,m^-(\Theta)|_{l>1}]=0.
 \eeq{13}
 On the other hand, the transformation law of $m^-$ can be found by
 performing a Lorentz transformation on the asymptotic 
 metric~\eqref{1}. It is given by eq.~\eqref{1b}  with $w=3/2$, from which it is
 obvious that neither boosts nor rotations vanish on $m_{l>1}$.
 
 We must then conclude that
  while the supertranslation invariant Lorentz generators are useful quantum operators -- in fact essential for defining unambiguously
   the angular momentum of the vacuum as well as other quantum numbers-- they should not be used to generate Lorentz 
   transformations on the fields.
   In fact, the supertranslation-invariant Lorentz charges
   are elements of the
   {\em universal enveloping algebra} of an enlarged BMS algebra that includes logarithmic
   supertranslations. Their explicit form is given in eq.~(9.7) of
   ref.~\cite{henn}.
   
   In this letter we defined the covariantization of the charge $J^{CWWY}_Y$;  in a future work we will expand on the results
    presented here and discuss the covariantization and interpretation of other charges, such as those
   given in~\cite{jkp,jp}.

   \subsection*{Acknowledgements} 

   We thank Gabriele Veneziano, Massimiliano Riva and Geoffrey Comp\`ere for useful discussions and one of the referees for pointing out to us
   ref.~\cite{winicour}. M.P. is supported in part by NSF grant PHY-2210349 and by the Leverhulme Trust through a Leverhulme Visiting Professorship at
 Imperial College, London.
 

\end{document}